\documentclass[%
 reprint,
superscriptaddress,
 amsmath,amssymb,
 aps, prb
]{revtex4-2}
\pdfoutput=1
\usepackage{hyperref}
\usepackage{graphicx}
\usepackage{dcolumn}
\usepackage{bm}
\usepackage{xcolor} 
\usepackage{orcidlink}
\usepackage{url}
\definecolor{color1}{rgb}{0,0.25,0.70}
\hypersetup{colorlinks=true, 
	linkcolor={color1},
	citecolor={color1}, 
	urlcolor={color1}
}

\begin{document}

\preprint{APS/123-QED}

\title{Consistency between the Green-Kubo formula and Lorentz model for predicting the infrared dielectric function of polar materials}

\author{Wei-Zhe Yuan~\orcidlink{0009-0001-4678-4344}}
\affiliation{School of Energy Science and Engineering, Harbin Institute of Technology, Harbin 150001, China}
\affiliation{Key Laboratory of Aerospace Thermophysics, Ministry of Industry and Information Technology, Harbin 150001, China}
\author{Yangyu Guo~\orcidlink{0000-0003-2862-896X}}
\affiliation{School of Energy Science and Engineering, Harbin Institute of Technology, Harbin 150001, China}
\affiliation{Key Laboratory of Aerospace Thermophysics, Ministry of Industry and Information Technology, Harbin 150001, China}
\author{Hong-Liang Yi~\orcidlink{0000-0002-5244-7117}}
\email[]{yihongliang@hit.edu.cn}
\affiliation{School of Energy Science and Engineering, Harbin Institute of Technology, Harbin 150001, China}
\affiliation{Key Laboratory of Aerospace Thermophysics, Ministry of Industry and Information Technology, Harbin 150001, China}
\date{\today}

\begin{abstract}
Accurate prediction of infrared dielectric functions in polar materials is fundamental for thermal and photonic applications,  yet it remains unexplored whether the two main methods, Green-Kubo formula and Lorentz model, can give unified predictions. In this work, we present a detailed comparison of these two approaches using MgO and LiH as prototypical cases employing both empirical rigid ion model~(RIM) and machine learning potential~(MLP). 
We demonstrate that the conventional Lorentz model fails to capture the multi-phonon absorption inherent in Green-Kubo method, which can be resolved via using the phonon self-energy as a generalization of the usual linewidth.
In addition, with RIM, a correction factor of $\varepsilon_\infty$ is required in the ionic contribution to infrared response to account for the electronic polarization effect, which is yet captured by MLP using the Born effective charges for calculating dipole moment.
The present benchmark study thus enables cross-validation of dielectric function calculations while providing mechanistic insights into the polarization dynamics.
\end{abstract}

\maketitle

\section{Introduction}
The infrared dielectric function serves as a fundamental thermophysical property in radiative heat transfer for aerospace and solar energy applications~\cite{RevModPhys.39.432,howell2020thermal,modest2021radiative,howell2023thermal,Zhao_2024,wu2022NatRevPhy,2019-nanophoto}. Significant research efforts have been dedicated to characterizing its temperature dependence, driven by operational requirements in extreme thermal environments~\cite{Tong2020PRB,Zhou2024PRL}. Nevertheless, experimental determination of high-temperature dielectric responses faces formidable technical challenges, including material degradation through oxidation and self-radiation~\cite{palik1998handbook}. These limitations underscore the critical need for theoretical frameworks capable of predicting temperature-dependent dielectric functions with first-principles accuracy.

The Lorentz model provides the foundational framework connecting the infrared dielectric responses to lattice dynamics in polar crystals~\cite{born1996dynamical}. Recent advances in first-principles perturbation theory have enabled \textit{ab initio} prediction of temperature-dependent phonon properties and associated dielectric functions~\cite{Tong2020PRB,2023Han_PRB,ZhouHao2024PRM,Tiwari2024PRB,PhysRevB.110.165412,Guozq2024APL}.  Nevertheless, 
the breakdown of quasiparticle approximation at elevated temperatures fundamentally challenges the validity of perturbative approach~\cite{PhysRevX.12.041011}.

MD simulation provides another way to compute the phonon properties required by the Lorentz model via spectral energy density (SED) analysis~\cite{BAO20121683}. It offers distinct advantages by inherently capturing all-order anharmonic interactions. On the other hand, with the Green-Kubo formula, MD provides a nonperturbative framework for computing infrared dielectric functions directly from equilibrium dipole moment fluctuation statistics~\cite{PhysRev.123.777,Gangemi_2015,DOMINGUES2018220,Chen2021JAP,PhysRevB.108.085434}. 
However, the consistency between these two fundamentally different approaches (i.e. Green-Kubo as a microscopic statistical method and Lorentz model as a phenomenological one) remains still ambiguous.
In principle, the infrared dielectric functions computed from these two methodologies should exhibit quantitative agreement when phonons manifest as well-defined quasiparticles, yet there are two inherent obstacles that limit the two approaches to give unified prediction. 

First, conventional calculation of infrared dielectric functions predominantly relies on the empirical RIM~\cite{1974RIM}, which inherently neglects electronic polarizability~\cite{dpsvol1}.
Actually, the interionic Pauli repulsion induces electronic charge deformation and relative displacement between electron clouds and nuclei~\cite{dpsvol1}, generating additional dipole moments that exceed rigid-charge predictions~\cite{Szigeti1949,Szigeti1950}. This dual nature of infrared polarization, arising from coupled electronic and ionic contributions~\cite{1991pssa,sirdeshmukh2007fifty}, presents fundamental challenges in simulation. Recently, Domingues et al.~\cite{DOMINGUES2018220} highlighted that the dielectric function calculated from the Green-Kubo formula with RIM provides only the term $\varepsilon(\omega)/\varepsilon_\infty$ as expressed in Lorentz model, while the underlying mechanism remains poorly understood. Furthermore, it remains to be investigated whether the Green-Kubo MD with more advanced MLP is capable of producing predictions consistent with the Lorentz model. 

Second, Lorentz model with frequency-independent phonon linewidth is equivalent to fitting the dipole moment correlation function with a single exponentially decaying oscillator of the form $e^{-t/\tau}\cos \left( 2\pi \omega _0t \right)$~\cite{Carati16,Chen2021JAP} ($\tau$: phonon lifetime; $\omega_{0}$: phonon frequency). However, this approach neglects numerous fine features of the correlation function that may be associated with multiphonon infrared absorption~\cite{1970PRSLA,2004AS,Sun2008PRB}, which is instead included naturally in Green-Kubo MD. 
Although the more generalized phonon self-energy obtained from perturbation theory is incorporated in Lorentz model to describe the multiphonon absorption~\cite{Sun2008PRB,Fugallo2018PRB,2022PRM,TiO2_2023PRB}, no work has been reported whether it is consistent with Green-Kubo MD to our best knowledge.
We will show in this work that the Green-Kubo formula reconciles the Lorentz model parameterized with phonon self-energy extracted from MD simulation.

In this work, we present a comprehensive comparison of the Green-Kubo formula and Lorentz model with the same interatomic potential for predicting infrared optical properties. We select two representative materials: lithium hydride~(LiH) with pronounced multi-phonon infrared absorption and magnesium oxide~(MgO) with less phonon anharmonicity. 
For each material, we employ RIM and MLP to assess the contribution of electronic polarization to infrared response. 
Additionally, we parameterize the Lorentz model using either phonon linewidth or phonon self-energy to investigate the impact of multiphonon absorption on the unification of the two approaches. Finally, we examine the potential nuclear quantum effects on the infrared dielectric function. 
The remaining of this work is organized as follows: The
methodology and simulation detals will be introduced in Sec.~\ref{sec2}, followed by a discussion of the results in Sec.~\ref{sec3}, and the concluding remarks will be made in Sec.~\ref{sec4}.
\section{METHODOLOGY AND SIMULATION DETAILS}\label{sec2}
In this section, we systematically outline the methodology used to predict the infrared optical properties of polar materials. We begin by introducing the Green-Kubo formula in Sec.~\ref{sec2A}. Subsequently, in Sec.~\ref{sec2B}, we provide a detailed summary of the Lorentz model parameterized by either phonon linewidth from SED or the phonon self energy within MD. Next, we analyse the effect of electronic polarization in MD in Sec.~\ref{sec2C}. Last, we give the details of MD simulation including the fitting procedure for RIM and the training procedure for MLP in Sec.~\ref{sec2D}.
\subsection{Green-Kubo formula}\label{sec2A}
The local dielectric function is related to the dielectric
susceptibility $\chi_{\alpha \beta}(\omega)$ as $\varepsilon _{\mathrm{GK}, \alpha \beta}(\omega )=\delta _{\alpha \beta}+\chi _{\alpha \beta}\left( \omega \right)$~\cite{PhysRev.123.777,Gangemi_2015}, $\omega$ is the photon frequency, $\delta_{\alpha,\beta}$ is the Kronecker delta, and the imaginary part of susceptibility $\chi''_{\alpha \beta}$ is calculated via the fluctuation-dissipation theorem (i.e. Green-Kubo formula)~\cite{PhysRev.123.777,Gangemi_2015,DOMINGUES2018220,Chen2021JAP,PhysRevB.108.085434}:
\begin{subequations}
	\label{eq:chi}
\begin{equation}
	\begin{aligned}	
		\chi''_{\alpha \beta}\left( \omega \right) =&\mathrm{Im}\dfrac{V}{\varepsilon _0k_{\mathrm{B}}T}\bigg[ \left< P_{\alpha}\left( 0 \right) \cdot P_{\beta}\left( 0 \right) \right> 
		\\
		&+i\omega \int_0^{\infty}{\left< P_{\alpha}\left( 0 \right) \cdot P_{\beta}\left( t \right) \right> e^{i\omega t}dt} \bigg],
	\end{aligned}	
\end{equation}
\begin{equation}
	\begin{aligned}		
		\chi''_{\alpha \beta}\left( \omega \right)=&\mathrm{Im}\frac{1}{\varepsilon _0Vk_{\mathrm{B}}T\omega ^2}\bigg[ \left< J_{\alpha}\left( 0 \right) \cdot J_{\beta}\left( 0 \right) \right> 
\\
&+i\omega \int_0^{\infty}{\left< J_{\alpha}\left( 0 \right) \cdot J_{\beta}\left( t \right) \right> e^{i\omega t}dt} \bigg],
\end{aligned}	
\end{equation}
\end{subequations}
where $V$ is the system volume, with $\varepsilon_0$ the dielectric permittivity of the vacuum and $k_{\mathrm{B}}$ represents the Boltzmann constant. In Eq.(\ref{eq:chi}), the ionic polarization of the system is calculated as the density of dipole moment $\mathbf{P}\left( t \right)$ or current density $\mathbf{J}\left( t \right)$:
\begin{equation}\label{eq:P}
	\mathbf{P}\left( t \right) =\frac{\sum_{lb}{q_b\mathbf{u}_{lb}\left( t \right)}}{V},
\end{equation}
\begin{equation}
	\mathbf{J}\left( t \right) =\sum_{lb}{q_b\mathbf{v}_{lb}\left( t \right)},
\end{equation}
where $\mathbf{u}_{lb}=\mathbf{r}_{lb}-\mathbf{r}_{lb,0}$ is the atomic displacement with respect to its equilibrium
position $\mathbf{r}_{lb,0}$, with $\mathbf{v}_{lb}$ the atomic velocity and $q_{b}$ is the charge of the $b$-th basis atom. The real part of susceptibility $\chi'_{\alpha \beta}$ is then calculated using Kramer-Kronig relation~\cite{2014YangJY}
\begin{equation}
\chi '_{\alpha \beta}\left( \omega \right) =\frac{2}{\pi}\int_0^{\infty}{\frac{\omega '\chi ''_{\alpha \beta}\left( \omega ' \right)}{\omega '^2-\omega ^2}d\omega '}.
\end{equation}
\subsection{Lorentz model}\label{sec2B}
The temperature-dependent dielectric function of polar dielectric materials can be written in parameterized Lorentz model~\cite{PhysRevB.110.165412}:
\begin{equation}\label{eq:1}
\varepsilon \left( \omega ,T \right) =\varepsilon _{\infty}\left[ 1+\frac{\omega _{\mathrm{LO}}^{2}(T)-\omega _{\mathrm{TO}}^{2}(T)}{\omega _{\mathrm{TO}}^{2}(T)-\omega ^2-i\omega \tau ^{-1}\left( T \right)} \right] ,
\end{equation}
where $\tau^{-1}$ is the TO phonon linewidth at $\Gamma$ point; $\omega_{\mathrm{TO}}$ and $\omega_{\mathrm{LO}}$ represent the zone-centered transverse and longitudinal optical phonon frequencies, respectively. $\omega_{\mathrm{TO}}$, $\omega_{\mathrm{LO}}$and $\tau^{-1}$ can be predicted from SED analysis via MD~\cite{PhysRevB.81.081411}. 

In this work, we use the version of the frequency- and wave-vector dependent SED [$\Phi \left( \mathbf{k},\omega \right)$] without eigenvector~\cite{PhysRevB.81.081411,Feng2015jap}:
\begin{equation}\label{eq:2}
\begin{aligned}	
\Phi \left( \mathbf{k},\omega \right) &=\frac{1}{4\pi \tau _0}\sum_{\alpha ,b}^{3,n}{\frac{M_b}{N_c}\left| \sum_l^{N_{c}}{\int_0^{\tau _0}{\dot{u}_{\alpha}^{l,b}\left( t \right)e^ { i\mathbf{k}\cdot \mathbf{r}_{0}^{l}-i\omega t } \mathrm{d}t}} \right|}^2
\\	
&=\sum_{\alpha}^3{\sum_b^n{\left| \dot{Q}_{\alpha}^{b}\left( \mathbf{k},\omega \right) \right|^2}},
\end{aligned}	
\end{equation}
where $u_{\alpha}^{l,b}$ is the $\alpha$-th component of the displacement of the $b$-th basis atom in the $l$-th unit cell. $\mathbf{r}_{0}^{l}$ is the equilibrium position of the $l$-th unit cell, $n$ and $N_c$ are the total number of basis atoms and unit cells, respectively. $M_b$ is the mass of $b$-th basis atom,  $\tau_0$ is the total integration time which should be five times greater than maximum phonon lifetime to get a converged result~\cite{PhysRevB.81.081411}. Phonon frequencies and lifetime are determined by fitting the SED through a superposition of $3n$ Lorentzian functions:
\begin{equation}\label{eq:3}
\Phi \left( \mathbf{k},\omega \right) =\sum_{\nu}^{3n}{\frac{C_{\mathbf{k},\nu}}{\left( \omega -\omega _{\mathbf{k},\nu} \right) ^2+\Gamma _{\mathbf{k},\nu}^{2}},}
\end{equation}
where $\nu$ is the phonon branch, $C_{\mathbf{k},\nu}$ is a constant for a given mode. $\omega_{\mathbf{k},\nu}$ is the peak position, related to the phonon frequency. $\Gamma _{\mathbf{k},\nu}$ is the half width at half maximum~(HWHM) of the peak, related to the phonon scattering rate as $\tau _{\mathbf{k},\nu}^{-1}=2\Gamma _{\mathbf{k},\nu}$~\cite{PhysRev.128.2589,PhysRevB.34.5058}.

However, zone-center $\omega_{\mathrm{LO}}$ cannot be directly captured by MD simulations due to the periodic boundary condition~\cite{BAO20121683,monacelli2024}. Alternatively, the zone-center $\omega_{\mathrm{LO}}$ can be extrapolated from LO phonon frequencies of the adjacent k-points~\cite{BAO20121683} from SED or obtained from perturbation theory with non-analytic correction~\cite{PhysRevB.55.10355} from MD input.

Equation~\ref{eq:1} does not, however, include the multiphonon infrared absorption process~\cite{2004AS}. A more general form of $\varepsilon(\omega)$ is thus~\cite{Cowley01101963}:
\begin{equation}\label{eq:8}
\varepsilon \left( \omega ,T \right) =\varepsilon _{\infty}\left[ 1+\frac{\omega _{\mathrm{LO}}^{2}\left( T \right) -\omega _{\mathrm{TO}}^{2}\left( T \right)}{\omega _{\mathrm{TO}}^{2}\left( 0 \right) -\omega ^2+2\omega _{\mathrm{TO}}\left( 0 \right) \Pi \left( \omega \right)} \right] ,
\end{equation}
where $\Pi \left( \omega \right) =\Delta \left( \omega \right) -i\Gamma \left( \omega \right)$ is the frequency-dependent self-energy of the TO phonon, with $\Delta \left( \omega \right)$ the anharmonic shift of TO phonon frequnency and $\Gamma \left( \omega \right)$ the damping due to anharmonic phonon-phonon scattering. The phonon self-energy can be calculated by phonon Green's function from MD~\cite{dpsvol2,SunTao2010PRB,vdmft2022PRB,vdmft2024PRB,2024LuYongPRB}. In the classical limit, the retarded single phonon Green’s function can be connected with the classical correlation function of modal velocity by the Kubo transform~\cite{Kubo1957}:
\begin{equation}
G^{\mathrm{R}}\left( \mathbf{k},\nu ,\omega \right) =\frac{\beta}{i\omega}\left< v_{\mathbf{k},\nu}^{*}(0)v_{\mathbf{k},\nu}(t) \right> _{\omega},
\end{equation}
where $\beta=1/k_{\mathrm{B}}T_{\mathbf{k},\nu}$. The phonon modal temperatures $T_{\mathbf{k},\nu}$ can be expressed as~\cite{Feng2017PRB}
\begin{equation}
T_{\mathbf{k},\nu}=\frac{1}{k_{\mathrm{B}}}\left< v_{\mathbf{k},\nu}^{*}\left( 0 \right) v_{\mathbf{k},\nu}\left( 0 \right) \right>.
\end{equation}
It is noted that the bracket $\left<  \right>_\omega $ represent the Fourier transform of the correlation function. The mode-projected velocity $v_{\mathbf{k},\nu}(t)$ is defined as
\begin{equation}
v_{\mathbf{k},\nu}\left( t \right) =\frac{1}{\sqrt{Nn}}\sum_l^N{\sum_b^n{\sqrt{M_b}}\mathbf{v}_{lb}\cdot}\mathbf{e}_{\mathbf{k},\nu}\exp \left( -i\mathbf{k}\cdot \mathbf{r}_{0}^{l} \right).
\end{equation}
From Dyson’s equation, $G^{\mathrm{R}}\left( \mathbf{k},\nu ,\omega \right)$ is represented in terms of phonon self-energy as
\begin{equation}\label{G_R}
G^{\mathrm{R}}\left( \mathbf{k},\nu ,\omega \right) =\frac{1}{\omega ^2-\omega _{\mathbf{k},\nu}^{2}-2\omega _{\mathbf{k},\nu}\Pi _{\mathbf{k},\nu}\left( \omega \right)}.
\end{equation}
Define $\Omega _{\mathbf{k},\nu}\left( \omega \right) =\omega _{\mathbf{k},\nu}+\Delta _{\mathbf{k},\nu}\left( \omega \right)$ as the renormalized phonon frequency, we have
\begin{equation}
G^{\mathrm{R}}\left( \mathbf{k},\nu ,\omega \right) \approx -\frac{1}{\Omega _{\mathbf{k},\nu}^{2}\left( \omega \right) -\omega ^2-2i\omega _{\mathbf{k},\nu}\Gamma _{\mathbf{k},\nu}\left( \omega \right)}.
\end{equation}
Therefore, $\Gamma _{\mathbf{k},\nu}\left( \omega \right)$ can be calculated from the real part $G^{\mathrm{R}'}$ and imaginary part $G^{\mathrm{R}''}$ as~\cite{1970PRSLA}
\begin{equation}
\Gamma _{\mathbf{k},\nu}\left( \omega \right) =-\frac{G^{\mathrm{R}''}\left( \mathbf{k},\nu ,\omega \right)}{2\omega _{\mathbf{k},\nu}\left\{ \left[ G^{\mathrm{R}'}\left( \mathbf{k},\nu ,\omega \right) \right] ^2+\left[ G^{\mathrm{R}''}\left( \mathbf{k},\nu ,\omega \right) \right] ^2 \right\}}.
\end{equation}
The real part of phonon self energy can be then obtained from Kramers–Kronig relations as~\cite{1967PRB}
\begin{equation}
\Delta _{\mathbf{k},\nu}\left( \omega \right) =-\frac{1}{\pi}\int_{-\infty}^{\infty}{\frac{\Gamma _{\mathbf{k},\nu}\left( \omega ' \right)}{\omega -\omega '}d\omega '}.
\end{equation}
\subsection{Electronic polarization effect in MD}\label{sec2C}
It is worth noting that MD based on RIM lacks the contribution of electronic polarization. To give a way to correct this polarization effect, we compare the difference between the phenomenological Lorentz formula in Eq.~\ref{eq:1} and one derived from Born-Huang theory~\cite{born1996dynamical} as below.

We consider a 1D polar diatomic chain. The masses of the positive and negative ions are given as $M_+$ and $M_-$, the displacements are $u_+$ and $u_-$ and the charges are $+q$ and $-q$. The following equations are then introduced~\cite{Huang1951}:
\begin{equation}\label{Eq:H1}
	\ddot{\mathbf{w}}=b_{11}\mathbf{w}+b_{12}\mathbf{E},
\end{equation}
\begin{equation}\label{Eq:H2}
	\mathbf{P}=b_{21}\mathbf{w}+b_{22}\mathbf{E},
\end{equation}
where $\mathbf{w}=\sqrt{\frac{M}{\Omega}}\left( \mathbf{u}_+-\mathbf{u}_- \right)$, $\mathbf{E}$ is electric field, $M=M_+M_-/\left( M_++M_- \right) $ is the reduced mass and $\Omega$ is the volume of a unit cell. $b_{11}$, $b_{12}$, $b_{21}$, $b_{22}$ are coefficients determined by polarization. Taking the Fourier transform of Eq.(\ref{Eq:H1}), we obtain:
\begin{equation}\label{Eq:H1w}
	-\omega ^2\mathbf{w}=b_{11}\mathbf{w}+b_{12}\mathbf{E}.
\end{equation}
Combing Eqs.(\ref{Eq:H2}) and (\ref{Eq:H1w}) and the definition of the dielectric displacement: $\mathbf{D}=\mathbf{E}+4\pi \mathbf{P}=\varepsilon \mathbf{E}$, we obtain the dielectric function:
\begin{equation}\label{Eq:epsH}
	\varepsilon =1+4\pi b_{22}+\frac{4\pi b_{12}b_{21}}{-b_{11}-\omega ^2}
\end{equation} 

When only ionic polarization is considered in MD with RIM, the density of dipole moment is:
\begin{equation}\label{Eq:P}
	\mathbf{P}=\dfrac{q}{\Omega}\left( \mathbf{u}_+-\mathbf{u}_- \right) =\dfrac{q}{\sqrt{M\Omega}}\mathbf{w}.
\end{equation}

Based on Newton's law, the motions of the positive and negative ions are:
\begin{equation}\label{Eq:M+}
	M_+\ddot{\mathbf{u}}_+=K \left( \mathbf{u}_--\mathbf{u}_+ \right) -K \left( \mathbf{u}_+-\mathbf{u}_- \right) +q\mathbf{E}_{\mathrm{eff}},
\end{equation}
\begin{equation}\label{Eq:M-}
	M_-\ddot{\mathbf{u}}_-=K \left( \mathbf{u}_+-\mathbf{u}_- \right) -K \left( \mathbf{u}_--\mathbf{u}_+ \right) -q\mathbf{E}_{\mathrm{eff}},
\end{equation}
where $K$ is the spring constant and $\mathbf{E}_{\mathrm{eff}}=\mathbf{E}+\frac{\mathbf{P}}{3\varepsilon _0}$ is the effective electric field. Combining Eq.(\ref{Eq:M+}) and Eq.(\ref{Eq:M-}) gives:
\begin{equation}\label{Eq:w}
	\ddot{\mathbf{w}}=-\frac{2K}{M}\mathbf{w}+\frac{q}{\sqrt{M\Omega}}\left( \mathbf{E}+\frac{\mathbf{P}}{3\varepsilon _0} \right).
\end{equation}
Compared Eq.(\ref{Eq:P}), Eq.(\ref{Eq:w}) to Eqs.(\ref{Eq:H1}-\ref{Eq:H2}), we determine the coefficients as:
\begin{equation}
	\begin{cases}
		b_{11}=-\dfrac{2K}{M}+\dfrac{q^2}{3\varepsilon _0M\Omega}\\
		b_{12}=b_{21}=\dfrac{q}{\sqrt{M\Omega}}\\
		b_{22}=0.\\
	\end{cases}
\end{equation}
Inserting these coefficient into Eq.(\ref{Eq:epsH}), we obtain the expression of local dielectric function:
\begin{equation}\label{Eq:eps}
	\varepsilon _{\mathrm{ion}}(\omega)=1+\dfrac{4\pi}{\Omega}\dfrac{\dfrac{q}{\sqrt{M}}\dfrac{q}{\sqrt{M}}}{\omega _{0}^{2}-\omega ^2},
\end{equation}
where $\omega _{0}^{2}=-b_{11}$. The expression in Eq.~\ref{Eq:eps} is similiar to the equation deduced from 3D lattice dynamic theory~\cite{dpsvol1,PhysRevB.55.10355}. When $\omega\rightarrow0$, $\varepsilon _{\mathrm{ion}}\left( 0 \right) =1+4\pi q^2/M\Omega \omega _{0}^{2}$.
In the other limit of $\omega\rightarrow\infty$, $\varepsilon _{\mathrm{ion}}\left( \infty \right) =1$. Thus, Eq.(\ref{Eq:eps}) will be:
\begin{equation}
	\varepsilon _{\mathrm{ion}}\left( \omega \right) =1+\frac{\left[ \varepsilon _{\mathrm{ion}}\left( 0 \right) -1 \right] \omega _{0}^{2}}{\omega _{0}^{2}-\omega ^2}.
\end{equation}
Considering the damping due to phonon scattering and the Lyddane–Sachs–Teller~(LST) relationship, we obtain the Lorentz formula without electronic degree of freedom:
\begin{equation}\label{Eq:eps_ion}
	\varepsilon _{\mathrm{ion}}\left( \omega \right) =1+\frac{\omega _{\mathrm{LO}}^{2}-\omega _{\mathrm{TO}}^{2}}{\omega _{\mathrm{TO}}^{2}-\omega ^2-i\omega \tau ^{-1}}.
\end{equation}
Compared Eq.(\ref{Eq:eps_ion}) with Eq.(\ref{eq:1}), we can deduce the correction term to account for the electronic contribution in MD simulation based on RIM:
\begin{equation}\label{Eq.corr}
	\varepsilon \left( \omega \right) =\varepsilon _{\infty}\varepsilon _{\mathrm{ion}}\left( \omega \right). 
\end{equation}

It is seen that the electronic polarization contributes not only to the usual term $\varepsilon_\infty$ but also to the infrared response of ions. This indicates that existing correction of dielectric function obtained via RIM-based MD by just adding $\varepsilon_\infty-1$~\cite{Gangemi_2015,Chen2021JAP,PhysRevB.108.085434} will much underestimate the result, as to be shown later in Sec~\ref{sec3A}. Also we will show that MD with MLP automatically include the electronic contribution to ionic infrared response.

\subsection{MD simulation details}\label{sec2D}
The Green-Kubo calculation, SED analysis and phonon self-energy calculation are done based on equilibrium molecular dynamics~(EMD) as implemented in large-scale atomic/molecular massively parallel simulator (\textsc{LAMMPS}) package~\cite{LAMMPS} with RIM and Graphics Processing Units Molecular Dynamics~(\textsc{GPUMD}) package~\cite{GPUMD2022} with neuroevolution potential~(NEP). The development of interatomic potential will be described below. During the EMD simulation, a $10\times10\times10$ cubic supercell of 8000 atoms with a time step of 0.5 fs is adopted. The size of the supercell has been tested to be sufficiently capture the long-range interaction well. In addition, periodic boundary conditions are imposed along the three directions of the system. The particle-particle particle-mesh method~\cite{EASTWOOD1980215} is implemented for the treatment of long-range Coulomb force of RIM with a cutoff radius of 10 \AA~and accuracy of $10^{-4}$ for the direct interactions in real space. First, $2\times10^{5}$ time steps are run under the \textit{NPT}~(isothermal-isobaric) ensemble for structure relaxation. Then, another $1\times10^{6}$ time steps are run under the \textit{NVE}~(microcanonical) ensemble. The trajectory of the system is output once per 20 time
steps during the \textit{NVE} run. It is worth noting that the atomic dynamics in above simulations are essentially classical. To check the influence of quantum dynamics~\cite{2019PRM,2024APL-classic} on infrared optical properties, we also perform a thermostatted ring-polymer molecular dynamics~(TRPMD) simulation as implemented in the path integral molecular dynamics~(PIMD) module of the \textsc{GPUMD} package~\cite{ying2024pimd}. In PIMD simulation, 40 beads are employed in all simulations which is sufficient to achieve full convergence~\cite{ying2024pimd}.
\subsubsection{Rigid ion empirical potential}
\begin{figure*}[htbp]
	\centering
	\includegraphics[width=0.9\linewidth]{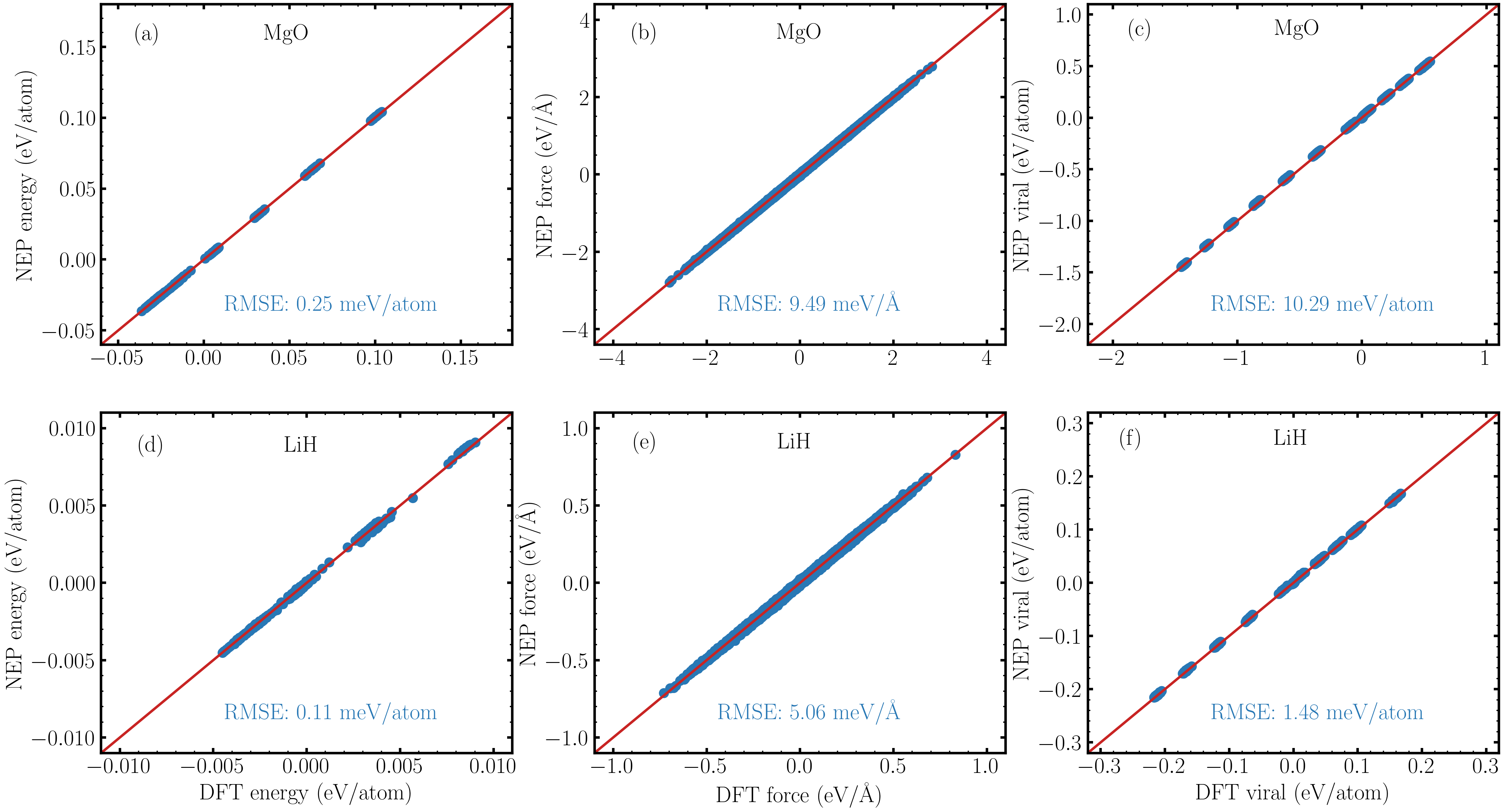}
	\caption{The partial plots of total energy, atomic forces, virial for MgO~(a-c) and LiH~(d-f). The insets show the RMSE of the testing datasets.}
	\label{fig:train}
\end{figure*}
We use a pairwise van Beest,
Kramer, and van Santen (BKS) potential form $\phi_{ij}$ between atoms $i$ and $j$ composed of a short-range Buckingham term and a long-range Coulomb interaction which can be written as~\cite{BKS}
\begin{equation}\label{eq:4}
\phi _{ij}=\frac{q_iq_j}{r_{ij}}+A\exp \left( -\frac{r_{ij}}{\rho} \right) -\frac{C}{r_{ij}^{6}},
\end{equation}
where $r_{ij}$ is the distance between atom $i$ and atom $j$, $q_i$ and $q_j$ are the partial charges of the atoms. The other atomic parameters for MgO are given in Table~\ref{tab1}~\cite{MgO1989Born}. While there are no BKS potential parameters for LiH in the literature, we choose to fit such a potential by general utility lattice program (\textsc{GULP})~\cite{gulp}. 
\begin{table}[htbp]
\caption{\label{tab1} Parameters in BKS potential of MgO~\cite{MgO1989Born}.}
\begin{ruledtabular}
\begin{tabular}{ccccc}
Pair of atoms &$q$ (e)& $A$ (eV) & $\rho$ (\AA) & $C$ (eV$\cdot$\AA$^6$) \\ \hline
	Mg-Mg  &$+1.4$& 1309363.25 & 0.104 & 0 \\ 
	O-O  &$-1.4$&2145.86142 & 0.300 & 30.22 \\ 
Mg-O  & &2543184.81 & 0.104 & 0 \\	
\end{tabular}
\end{ruledtabular}
\end{table}

The BKS potential for LiH is developed by fitting the phonon dispersion spectrum from density functional theory (DFT)-based first-principles calculations with its details given in the development of MLP below. The total number of parameters allowed to be fitted during optimization is 11, including 9 potential parameters for three different types of interactions and 2 partial charges. The so-called relax fitting methodology is employed~\cite{Gale01011996}, whereby the structure is fully optimized at each evaluation of the fitting function. The optimized potential parameters are shown in Table~\ref{tab2}. The partial charge agrees with the Szigeti effective charge from the experiment~(LiH: $0.52\pm0.02$)~\cite{LiH1967}. To further validate its accuracy, we computed the ground-state (0 K) phonon dispersion and compare it with DFT and experiment in Figs.~\ref{fig:mgodis} and \ref{fig:lihdis} of Appendix~\ref{sec:a1}, while the anharmonic properties are tested in subsequent MD simulations. 
\begin{table}[htbp]
	\caption{\label{tab2} Parameters in BKS potential of LiH.}
	\begin{ruledtabular}
		\begin{tabular}{ccccc}
			Pair of atoms &$q$ (e)& $A$ (eV) & $\rho$ (\AA) & $C$ (eV$\cdot$\AA$^6$) \\ \hline
			Li-Li  &$+0.46$& 63004.568 & 0.182 & 0.0438 \\ 
			H-H  &$-0.46$&14301.840 & 0.208 & 26.288 \\ 
			Li-H  & & 10600.001 & 0.156 & 0.0019 \\ 
		\end{tabular}
	\end{ruledtabular}
\end{table}

According to Lorentz-Lorenz relation~\cite{1979pssb}, for unpolarizable rigid ions $\varepsilon_{\infty}=1$, while $\varepsilon_{\infty}$ from experiment~(MgO: 3.01~\cite{palik1998handbook}; LiH: 3.61~\cite{LiH1967}.) should be used to correct the electronic polarization effect on dielectric function as discussed in Sec.~\ref{sec2C}. 
\subsubsection{Machine learning potential}
We use the NEP approach~\cite{NEP2021,GPUMD2022} to construct accurate MLP models of MgO and LiH. A total of 1000 structures were sampled for MgO and LiH, with different lattice expansion respect to ground state~(-2\% $\sim$ +5\%) and random displacement with magnitude of 0.1 \AA. Each structure contains 128 atoms. The \textsc{Quantum Espresso} package~\cite{2020QE} with the projector-augmented wave method is used to obtain the energy, forces, and virial of each structure. In the DFT calculations, the Perdew-Burke-Ernzerhof revised for solids (PBESOL) function is used to describe the exchange-correlation of electrons~\cite{2009PRB_PBESOL}. The kinetic energy cutoff for wave function and charge density have been selected separately: 60 and 300 Ry for MgO; 70 and 500 Ry for LiH. Convergence threshold for self consistency was $10^{-8}$ eV. The $\mathbf{k}$-point mesh was set to Gamma only. Then, 80\% and 20\% structures were randomly selected to form the training and testing datasets, respectively. 

During the training processes, the radial and angular descriptor components were set to a cutoff radius of 5~\AA. There may be no advantage in increasing the cut-off radius further, as discussed in Ref.~\cite{deng2019electrostatic} and Appendix~\ref{sec:a2}. The parity plots and accuracy metrics are shown in Fig.~\ref{fig:train}. The root mean square error (RMSE) values of the total
energy, atomic forces, and virial for the testing dataset
are: 0.25 meV/atom, 9.49 meV/\AA~and 10.29 meV/atom for MgO; 0.11 meV/atom, 5.06 meV/\AA~and 1.48 meV/atom for LiH. The training datasets and the trained NEP models for MgO and LiH are freely available~\cite{Yuan}.
To further validate the NEP accuracy, we computed the ground-state (0 K) phonon dispersion using the finite displacement method as shown in Figs.~\ref{fig:mgodis} and \ref{fig:lihdis} of Appendix~\ref{sec:a1}. 
\begin{figure*}[htbp]
	\centering
	\includegraphics[width=0.9\linewidth]{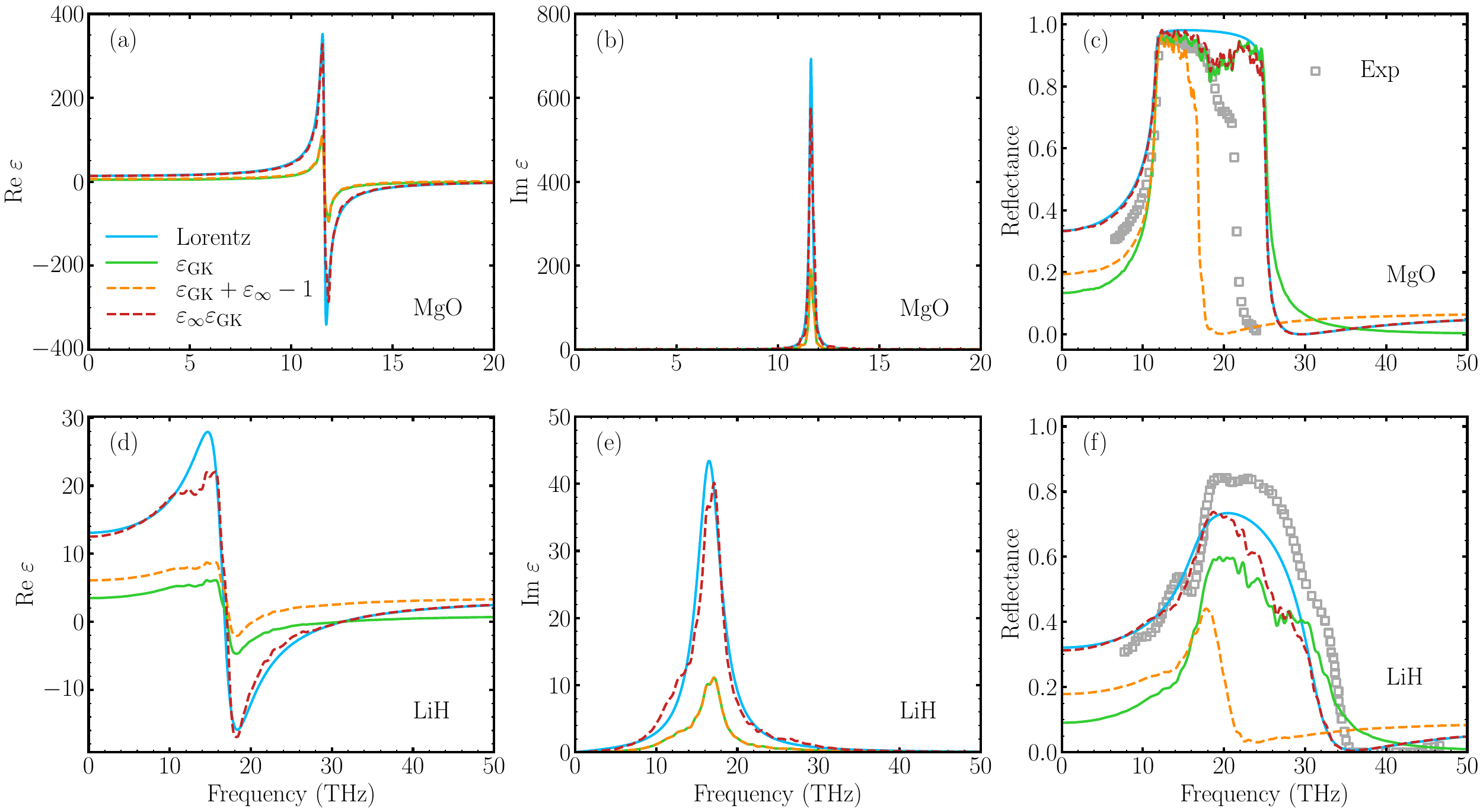}
	\caption{The real and imaginary part of infrared dielectric function and semi-infinite normal reflectance of MgO~(a)-(c) and LiH~(d)-(f) calculated from the Green-Kubo~(GK) formula and SED parameterized Lorentz model based on RIM potential at 300 K. Reflectance measurements are taken from Ref.~\cite{1966PR,LiH1967} (grey squares). Blue solid lines are calculated from SED parameterized Lorentz model; green solid lines are calculated from raw dielectric function $\varepsilon_{\mathrm{GK}}$ from the Green-Kubo formula without any correction; orange and red dashed lines are calculated from $\varepsilon_{\mathrm{GK}}+\varepsilon_\infty-1$ and $\varepsilon_\infty\varepsilon_{\mathrm{GK}}$, respectively. }
	\label{fig:RIM}
\end{figure*}

Although MLP inherently captures electron polarization effects on atomic motion through learning the local environment-dependent charge redistribution in DFT, it has no apparent charge in its potential energy expression. To address this limitation, we assign Born effective charges from density functional perturbation theory~(DFPT)~(1.95 for MgO and 1.02 for LiH) to the atoms when calculating the dipole moment inspired by the narrow charge distributions in ionic systems~\cite{deng2019electrostatic}. We will then compare the predicted infrared optical properties with experiments to analyze the latent effect of charge fluctuations.
Finally, $\varepsilon_{\infty}$ calculated via DFPT yield values of 3.16 for MgO and 4.72 for LiH. These results overestimate the experimental references (MgO: 3.01~\cite{palik1998handbook}; LiH: 3.61~\cite{LiH1967}),  attributed to the inherent limitations of DFT. The influence of $\varepsilon_\infty$ and effective charge on reflectance is discussed in Appendix~\ref{sec:a3}. To ensure quantitative agreement with optical measurements, all subsequent analyses will adopt the experimentally determined $\varepsilon_{\infty}$ values unless explicitly stated otherwise.
\section{Result and discussion}\label{sec3}
In this section, we first show the predicted infrared optical properties by RIM and NEP in Sec.~\ref{sec3A} and Sec.~\ref{sec3B1} respectively. Comparing the results predicted by the Green-Kubo formula with Lorentz model, we find that RIM requires $\varepsilon_\infty$ correction, whereas MLP inherently capture electronic polarization but necessitate Born effective charges for explicit dipole moment quantification. 
Then, we demonstrate the role of phonon self energy to unify the Green-Kubo formula and Lorentz model for infrared dielectric function predictions in Sec.~\ref{sec3B}. We also compare the infrared optical properties from MD simulation and perturbation theory to analyze the strengths and weaknesses of each. Last, nuclear quantum effect on infrared dielectric function is analysed in Sec.~\ref{sec3D}. 
\subsection{Infrared optical properties by RIM}\label{sec3A}
In Fig.~\ref{fig:RIM}, we demonstrate different kinds of calculations of infrared optical properties~(complex dielectric function and reflectance) for MgO and LiH based on RIM. The semi-infinite normal reflectance $R$ can be determined with the relation
\begin{equation}
	R\left( \omega \right) =\left| \frac{\sqrt{\varepsilon \left( \omega \right)}-1}{\sqrt{\varepsilon \left( \omega \right)}+1} \right|^2.
\end{equation}
Usually, prior works~\cite{Gangemi_2015,Chen2021JAP,PhysRevB.108.085434} only considered the ultraviolet electronic polarization correction as $\varepsilon_{\mathrm{GK}}+\varepsilon_\infty-1$, with $\varepsilon_{\mathrm{GK}}$ the raw dielectric function from Green-Kubo formula. Nevertheless, electron polarization also contributes to the infrared response of ions. When both corrections are considered, the dielectric function is $\varepsilon_\infty\varepsilon_{\mathrm{GK}}$ as discussed in Sec.~\ref{sec2C}. 

For MgO, correcting the polarization in RIM using $\varepsilon_\infty\varepsilon_{\mathrm{RIM}}$ yields the dielectric function aligning closely with that obtained from the SED parameterized Lorentz model. Consequently, the reflectance also shows excellent agreement. However, if the electron polarization correction is omitted, the infrared dielectric function of MgO is significantly underestimated. This discrepancy leads to deviations in the reflectance outside the Reststrahlen region.  When the background ultraviolet dielectric constant is corrected in the conventional manner~\cite{Gangemi_2015,Chen2021JAP,PhysRevB.108.085434}, the real part of $\varepsilon(\omega)$ is slightly enhanced. However, it is evident that $\omega_{\mathrm{LO}}$~[$\mathrm{Re}\,\varepsilon \left( \omega _{\mathrm{LO}} \right) =0$]  is significantly underestimated, resulting in a noticeable bias in the reflectance as depicted in Fig.~\ref{fig:RIM}(c). To account for the effect of electron polarization on the RIM, it is essential to calculate the dipole moment using the effective charge. Alternatively, when employing partial charges, Eq.~\ref{Eq.corr} should be utilized. Furthermore, the SED parameterized Lorentz model fails to describe the small dip in the infrared reflectance within the Reststrahlen region~\cite{PhysRevB.99.220304} mainly due to the multi-phonon process is observed.

For LiH, similar conclusions can be drawn, with the notable exception that the multi-phonon absorption feature is more pronounced in the dielectric function due to its stronger anharmonicity. However, when comparing the reflectance predicted by RIM for both MgO and LiH, significant deviations from experimental results are observed. This discrepancy may arise from the inherent limitations of empirical potential models in accurately describing the complex dynamic properties of solids. Even fitting the BKS potential function of LiH to DFT data, the resulting model fails to capture the true anharmonic properties of the material. This shortcoming is likely due to the fundamental limitation of empirical two-body potentials, which neglect many-body interactions~\cite{2003manybody}. Thus, more reliable potential models are required to achieve accurate predictions of infrared optical properties.
\subsection{Infrared optical properties by MLP}\label{sec3B1}
\begin{figure*}[htbp]
	\centering
	\includegraphics[width=0.9\linewidth]{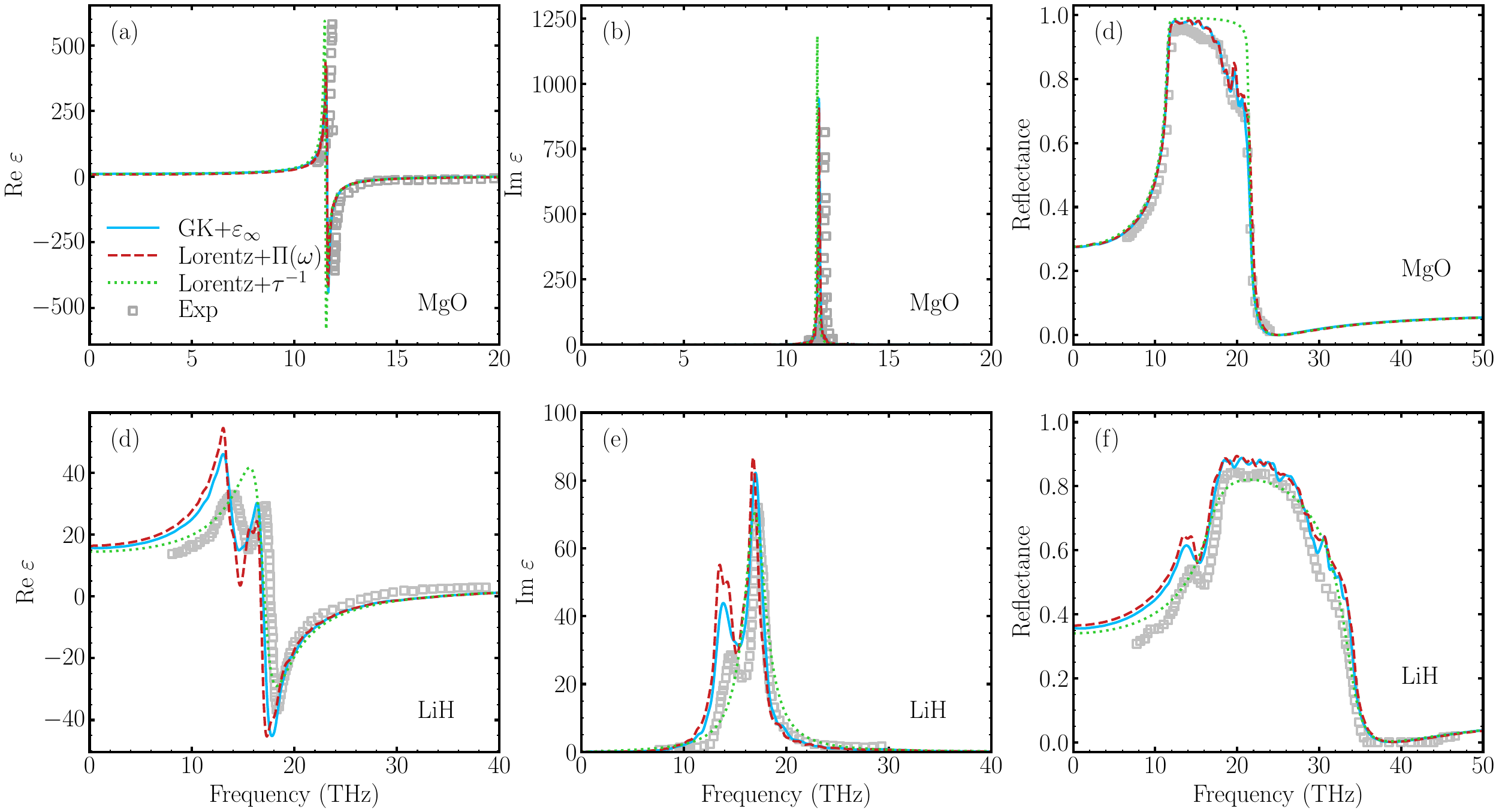}
	\caption{The real and imaginary part of infrared dielectric function and semi-infinite normal reflectance of MgO and LiH calculated from the Green-Kubo~(GK) formula ~(Blue solid lines) and parameterized Lorentz model with frequency dependent phonon self energy~(red dashed lines) or TO phonon scattering rate~(green dotted lines) based on NEP at 300 K as compared with experimental data~\cite{1966PR,2014YangJY,LiH1967} (grey squares). }
	\label{fig:nep}
\end{figure*}

In Fig.~\ref{fig:nep}, we present various kind of calculations of infrared optical properties~(complex dielectric function and reflectance) for MgO and LiH based on NEP. As previously discussed in Sec~\ref{sec2D}, we assign Born effective charges from DFPT~(1.95 for MgO and 1.02 for LiH) to the atoms when calculating the dipole moment. In addition, the force error in training NEP~\cite{Xiong2024JCP,Zhou2025MTP} has a negligible influence on predicting infrared optical properties of these materials whose anharmonicity are not very weak as discussed in Appendix~\ref{sec:error}.

For MgO, the dielectric function [Figs.~\ref{fig:nep}(a) and (b)] and reflectance [Fig.~\ref{fig:nep}(c)] predicted by the Green-Kubo formula with NEP exhibit excellent agreement with experimental measurements, except the slight underestimation of TO frequency attributed to the choice of pseudopotentials in DFT calculation. Notably, despite the limitations in accurately predicting LO mode near the $\Gamma$-point as evidenced in Fig.~\ref{fig:mgodis} from Appendix~\ref{sec:a1}, NEP has proven to be a reliable tool for predicting infrared absorption characteristics. This reliability stems from the fact that TO ($\Gamma$) is the sole one-phonon infrared active mode, while multi-phonon infrared absorption  predominantly occurs at critical points~(K, X, L, W), where the peculiar structure of phonon density of state facilitates such processes~\cite{JPC1973}. Importantly, both of them are not directly relevant to LO phonons near the Brillouin zone center.

For LiH, which exhibits more pronounced anharmonic phonon behavior, an additional prominent absorption peak appears below the TO one-phonon infrared-active peak at 13.9 THz~[Figs.~\ref{fig:nep}(d) and (e)]. This extra peak is attributed to anharmonic phonon-phonon interactions, as previously discussed in Sec~\ref{sec3A}. While the TO one-phonon infrared active peak is  well-predicted by NEP, the multi-phonon infrared absorption shows only quasi-quantitatively consistent with measurement~\cite{LiH1967}, with both absorption peaks slightly underestimated compared to experiment. Additionally, the static dielectric constant $\varepsilon(0)$=15.6 is larger than the measurement~(12.9~\cite{LiH1967}). This discrepancy may stem from the fact that DFPT overestimates Born effective charge relative to experiment. Consequently, this leads to an overestimation of $\omega_{\mathrm{LO}}$ and differences between the predicted and experimental reflectance outside the Reststrahlen region~[Fig.~\ref{fig:nep}~(f)]. In Appendix~\ref{sec:a3}, we recalculated the reflectance of LiH using NEP with experimental $\varepsilon_\infty$ and effective charge~\cite{LiH1967}. The resulting predicted reflectance shows better agreement with experiment. However, the calculations remain imperfect in the frequency range below TO, potentially due to underestimated TO phonon frequency or unaccounted effects such as long-range interactions~\cite{2022JCP} and nonlinear dipole moment effects~\cite{1970pssb}.

It is worth noting the partial charge in RIM is smaller than DFPT-calculated Born effective charge adopted in MLP. 
Actually, the Born effective charge $Z_{b,\beta \alpha}^{*}$ is defined as~\cite{PhysRevB.55.10355}:
\begin{equation}
Z_{b,\beta \alpha}^{*}=\Omega \left. \frac{\partial \mathcal{P} _{\mathrm{mac},\beta}}{\partial u_{\alpha}^{b}\left( \mathbf{k}=0 \right)} \right|_{\mathbf{E}=0},
\end{equation}
where the macroscopic polarization $\mathcal{P} _{\mathrm{mac}}$ incorporates contributions from both ionic displacements and electronic polarization. On the other hand, electron clouds mediate a screening effect that attenuates interionic Coulomb interactions by a factor of $\varepsilon_{\infty}^{-1}$~\cite{PhysRevB.1.910} as naturally included in MLP. In the framework of the RIM this screening could be mimicked through the use of partial charges~\cite{Salanne10102011}.
As a result, although the partial charge approximation improves Coulomb interaction accuracy, the RIM underestimates the ionic charge by a factor of $\varepsilon_{\infty}^{-0.5}$ and thus the dipole moment. Conversely, MLP is equivalent to using different charges in atomic dynamics and calculating dipole moments, so no correction is required. Furthermore, depending on the condition imposed on the macroscopic electric field, different concepts of charge have historically been introduced~\cite{PhysRevB.58.6224}. The underlying mechanisms for why Born effective charges enable accurate predictions of infrared optical responses, or whether alternative charge definitions could potentially achieve superior predictive capabilities, remain to be elucidated through more profound physical insights.

In summary, while NEP does not explicitly incorporate long-range interactions or charge deformation~\cite{chemrev}, it remains capable of predicting the primary features of infrared optical properties. It indicates the polarization induced by charge deformation is much less than that induced by ionic displacement in MgO and LiH.
\subsection{Unification of the Green-Kubo method and Lorentz model}\label{sec3B}
As discussed above, the SED parameterized Lorentz model fails to account for infrared multi-phonon absorption. To achieve consistency between the Lorentz model and the Green-Kubo formula, a frequency-dependent phonon self-energy is required~\cite{Cowley01101963}. In this work, we extracted the frequency-dependent phonon self-energy through the retarded Green's function via MD. To validate its accuracy, we compare the computed phonon self-energy with experimental results derived from reflection and transmission spectra~\cite{2004AS}, as shown in Fig.~\ref{fig:se} in Appendix~\ref{sec:a4}. Overall, the MD-derived phonon self-energy exhibits good agreement with experiment. The minor discrepancies observed may be attributed to inaccuracies in the LO phonon frequencies by NEP. The multi-phonon infrared absorption in MgO is primary caused by the lowest frequency peak caused by the phonon summation process. The probability of this process is proportional to $1+N_1+N_2$~\cite{PhysRevB.64.094301}, where $N_i$ is the occupation number of phonon mode $i$. Therefore, this summation mode is expected to maintain a non-vanishing intensity at low temperatures.

By utilizing the phonon self-energy of MgO and LiH extracted from MD, we calculate the dielectric function according to Eq.~\ref{eq:8} and derive the corresponding reflectance, as presented in Fig.~\ref{fig:nep}. The calculated results demonstrate excellent consistency with those predicted by the Green-Kubo formula. This strong agreement underscores the necessity of incorporating frequency-dependent phonon self-energy to bridge the Green-Kubo formula and Lorentz model.

\begin{figure}[htbp]
	\centering
	\includegraphics[width=1\linewidth]{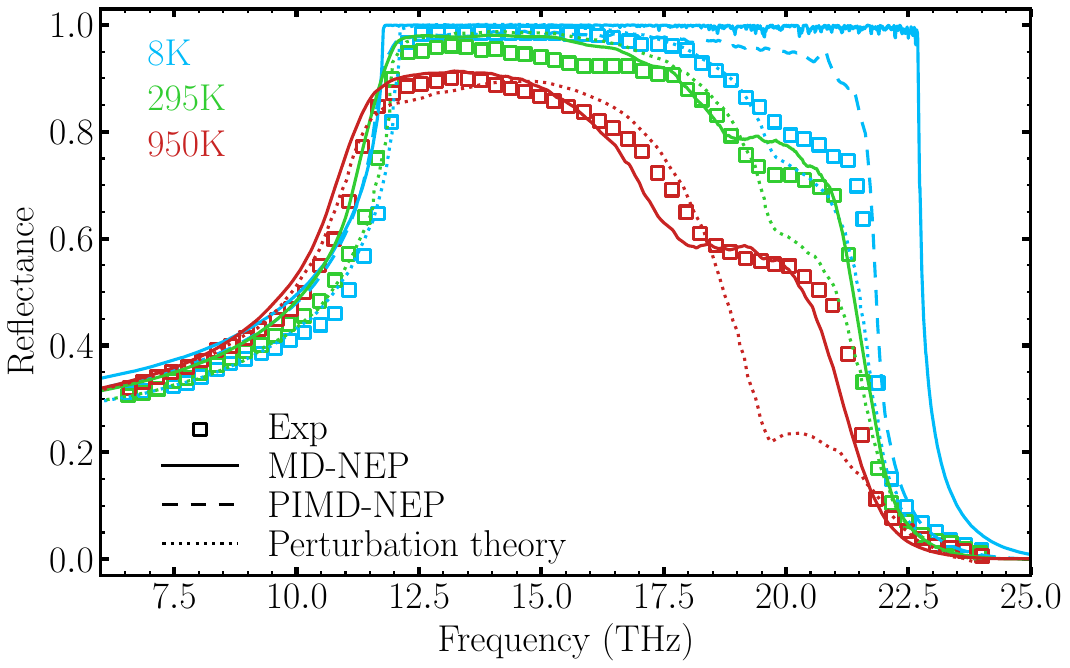}
	\caption{Infrared semi-infinite normal reflectance of MgO at different temperatures [8 K~(blue), 295 K~(green) and 950 K~(red)]: calculations from the Green-Kubo formula with NEP~(classic MD: solid lines; PIMD: dashed lines) versus measurements~\cite{1966PR}~(squares) and perturbation theory~\cite{Fugallo2018PRB}~(dotted lines). The reflectance at 5 K calculated by perturbation theory in the Ref.~\cite{Fugallo2018PRB} is approximately the same as that at 8 K.}
	\label{fig:rmgo}
\end{figure}

The phonon self-energy within the Lorentz model can also be determined using perturbation theory~(PT)~\cite{Sun2008PRB,Fugallo2018PRB}. Thus we further compare the infrared optical response predicted by MD and perturbation theory~\cite{Fugallo2018PRB}, as illustrated in Fig.~\ref{fig:rmgo}. The reflectance calculated by Green-Kubo and Lorentz model parameterized by self-energy via MD are very close regardless of temperature, thus we only show the results of former.

At 295~K and 950~K, the reflectance predicted by the Green-Kubo formula demonstrates better agreement with experiment compared to  PT above 18 THz and achieves similar prediction accuracy as PT below 18 THz. This discrepancy primarily arises from the fact that MD inherently accounts for all orders of anharmonic interactions, whereas PT typically considers anharmonic interactions up to the fourth order~\cite{Fugallo2018PRB}. Additionally, PT may lose accuracy at very high temperatures due to the breakdown of phonon quasiparticle picture. Conversely, at cryogenic temperature~(8 K), the reflectance spectra predicted by MD deviate significantly from the experimental one, which instead aligns much better with PT. This discrepancy could be probably attributed to the influence of isotopic disorder (negligible at higher temperature)~\cite{Fugallo2018PRB} as well as the quantum effect. PT offers a more straightforward approach for calculating isotopic-disorder scattering based on Tamura theory~\cite{1983Tamura}, whereas MD faces challenges in addressing this phenomenon due to limitations in simulation box size. Besides, quantum effects cannot be ignored at cryogenic temperatures as to be discussed in Sec.~\ref{sec3D}. Evidently, PIMD prediction considering nuclear quantum effects becomes closer to the experimental result. However, there is still some discrepancy which might stem from inherent limitations in the numerical implementation of PIMD~\cite{2014pimd} or limit simulation box size. 
In summary, PT is more suitable at cryogenic temperatures, while MD provides more accurate predictions at higher temperatures. 
\subsection{Nuclear quantum effect on infrared dielectric function}\label{sec3D}
In classical MD, zero-point energy is absent, and atomic motion decreases and approaches zero as the temperature decreases to absolute zero. In contrast, within the quantum mechanical framework, the energy does not vanish due to inherent vacuum fluctuations. For materials containing light elements, such as LiH, nuclear quantum effect (tunneling and zero-point motion) can play an exceptionally significant role. These effects have a profound impact on the material's physical properties~\cite{markland2018nuclear}.
\begin{figure}[htbp]
	\centering
	\includegraphics[width=1.0\linewidth]{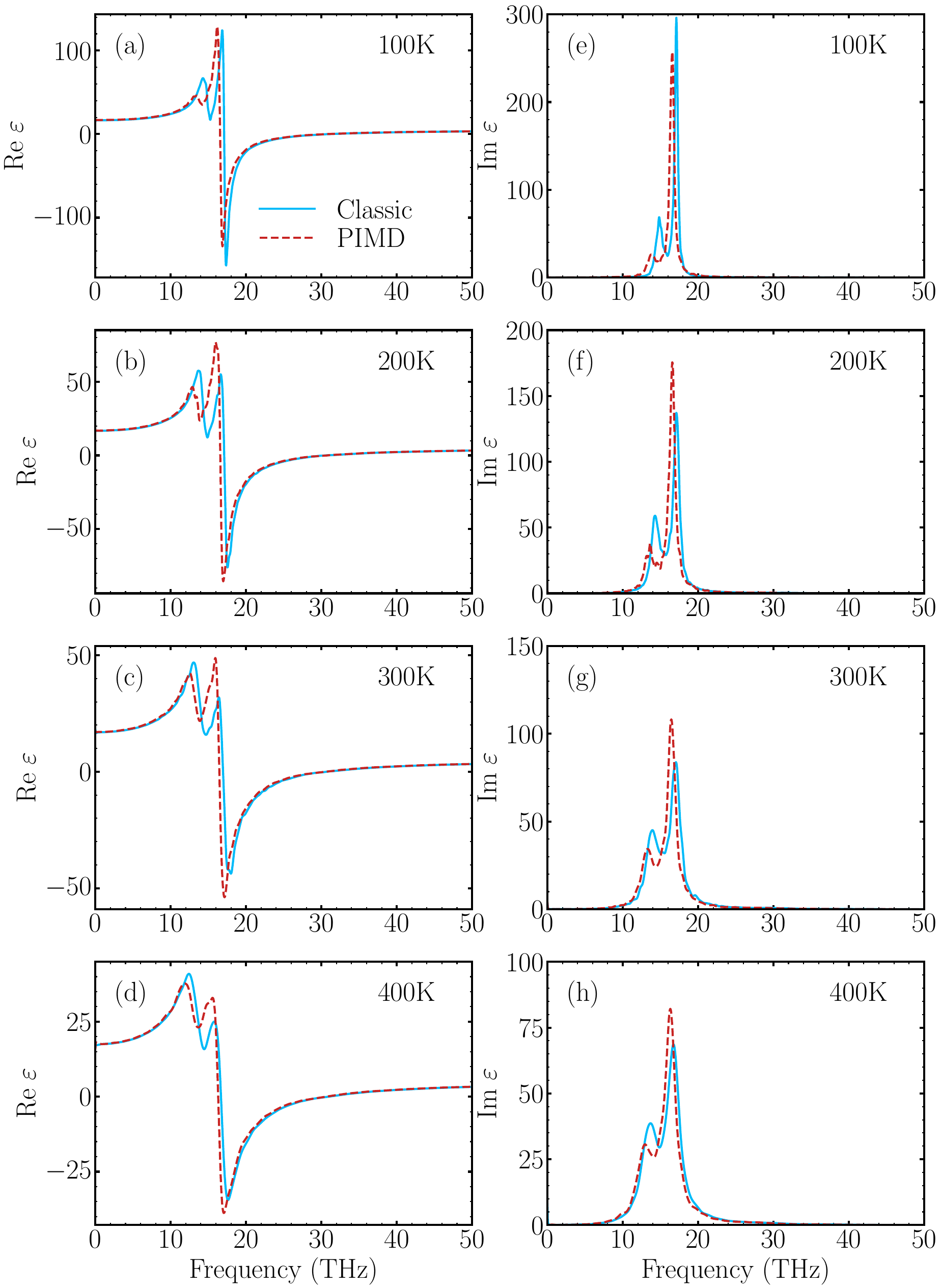}
	\caption{The temperature-dependent infrared dielectric function of LiH calculated from Classical MD~(blue solid lines) and PIMD~(red dashed lines) with NEP: (a)-(d) real part, (e)-(h) imaginary part.}
	\label{fig:liht}
\end{figure}

In Fig.~\ref{fig:liht}, we present a comparison of the infrared dielectric function of LiH computed using classical MD and PIMD both with NEP. As the temperature increases, nuclear quantum effect on the dielectric function is gradually reduced. Additionally, the optical phonon frequency predicted by PIMD is slightly lower than that obtained from classical MD. This discrepancy may arise from the additional zero-point energy in PIMD, which leads to an increased lattice constant and phonon softening. Notably, even at 400 K, differences in the predicted dielectric functions persist, highlighting the significant nuclear quantum effects in light-element systems. In contrast, for MgO, which has a heavier atomic mass, nuclear quantum effect become negligible at room temperature as demonstrated in Fig.~\ref{fig:nep3} in Appendix~\ref{sec:a5}. Therefore, careful consideration of nuclear quantum effects is essential when studying infrared optical properties of light-element materials or at cryogenic temperatures.
\section{Conclusion}\label{sec4}
In conclusion, we present a comprehensive comparison of the Green-Kubo formula and Lorentz model under the same interatomic potential (RIM and MLP) for predicting infrared optical properties.
We identify that two key factors that limit the Green-Kubo formula and Lorentz model to give unified predictions: electron polarization and multiphonon absorption. With RIM, a correction factor of $\varepsilon_\infty$ is required in the raw dielectric function to account for the electronic polarization. MLP inherently captures it in atomic dynamics through the Born effective charges for explicit dipole moment quantification. Furthermore, the failure of the conventional Lorentz model to resolve multi-phonon absorption is overcome by using MD-derived phonon self-energy, which thus establishes quantitative consistency with Green-Kubo predictions. Additionally, we demonstrate that classical MD with MLP efficiently models dielectric responses at elevated temperatures yet fails at cryogenic temperatures due to unresolved isotope scattering and nuclear quantum effects.  This work lays a solid foundation for accurately predicting temperature-dependent infrared optical properties of polar materials, particularly for thermal radiation and photonics applications.
\begin{acknowledgments}
This paper was supported by the National Natural Science Foundation of China (Grants No. U22A20210), the National Natural Science Foundation for Excellent Young Scientists Fund Program (Overseas) and starting-up funding from Harbin Institute of Technology~(Grant No. AUGA2160500923).
\end{acknowledgments}
\appendix
\section{The influence of NEP radius cutoff on dielectric function}\label{sec:a2}
\begin{figure}[htbp]
	\centering
	\includegraphics[width=1\linewidth]{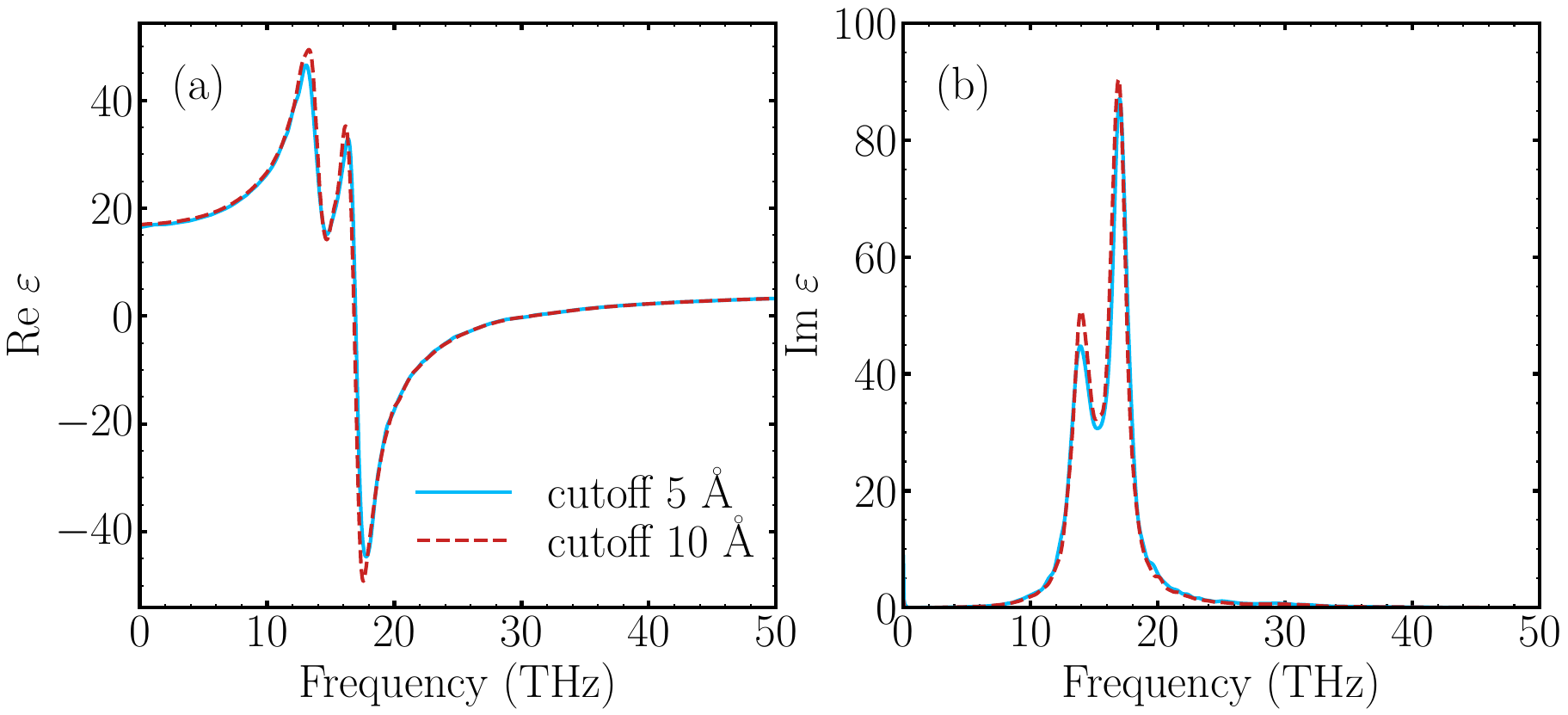}
	\caption{The infrared dielectric function of LiH calculated from NEP with cutoff radius of 5 \AA~(blue solid lines) and 10 \AA~(red dashed lines): (a) real part; (b) imaginary part.}
	\label{fig:cut}
\end{figure}

We developed an additional NEP for LiH employing a radius cutoff of 10 \AA. For the testing dataset, the root mean square error (RMSE) values for total
energy, atomic forces, and virial were 0.09 meV/atom, 5.46 meV/\AA~and 2.11 meV/atom, respectively. The infrared dielectric function predicted from it is presented in Fig.~\ref{fig:cut}. Comparatively, the RMSE and derived dielectric function exhibit only minor differences when contrasted with the potential functions utilizing a radius cutoff of 5 \AA. However, the computational cost increases 95\%, as demonstrated using a RTX4090D GPU.
\section{Phonon dispersion relations of MgO and LiH}\label{sec:a1}
\begin{figure}[htbp]
	\centering
	\includegraphics[width=0.7\linewidth]{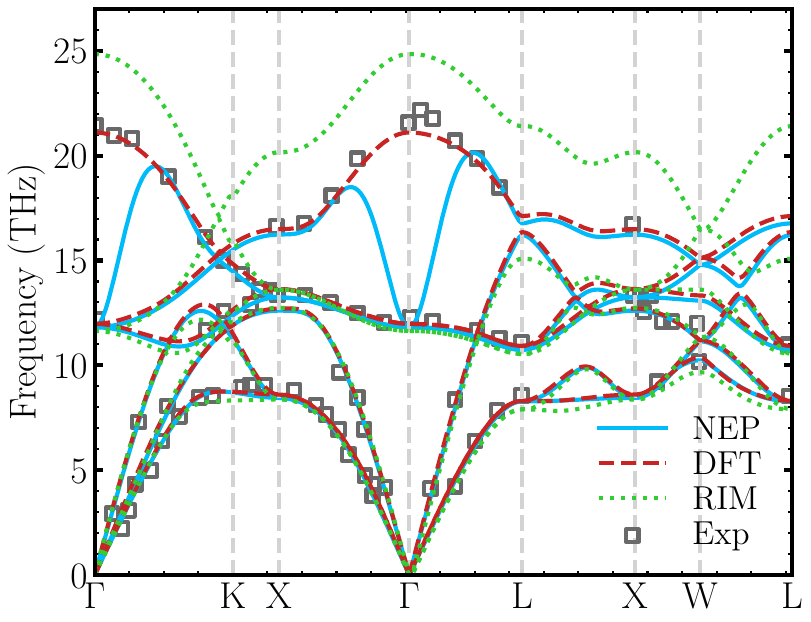}
	\caption{Phonon dispersion relations of MgO from: (a) NEP~(blue solid lines); (b) DFT~(red dashed lines); (c) RIM~(green dotted lines); (d) experiment~(grey squares)~\cite{2003JCP}.}
	\label{fig:mgodis}
\end{figure}

\begin{figure}[htbp]
	\centering
	\includegraphics[width=0.7\linewidth]{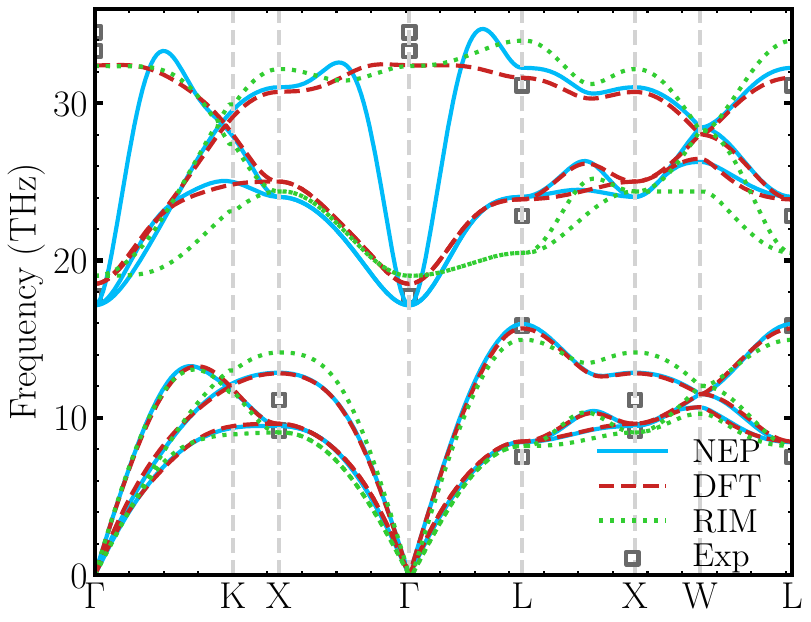}
	\caption{Phonon dispersion relations of LiH from: (a) NEP~(blue solid lines); (b) DFT~(red dashed lines); (c) RIM~(green dotted lines); (d) experiment~(grey squares)~\cite{ROMA1996203}.}
	\label{fig:lihdis}
\end{figure}

To further validate the accuracy of interatomic potential model, we employed the finite displacement method to compute the ground-state (0 K) phonon dispersion by RIM, NEP and DFT, as illustrated in Figs.~\ref{fig:mgodis} and \ref{fig:lihdis}. Overall, the phonon spectra derived from NEP, DFT and measurement show good agreement. However, discrepancies are observed in LO phonon frequencies near the $\Gamma$-point for NEP, which can be attributed to the absence of explicit long-range interactions in the model. In terms of performance, RIM demonstrates higher accuracy in predicting acoustic phonons compared to optical phonons.
\section{The influence of force error in training NEP on predicting infrared optical properties}\label{sec:error}
\begin{figure}[h!]
	\centering
	\includegraphics[width=1\linewidth]{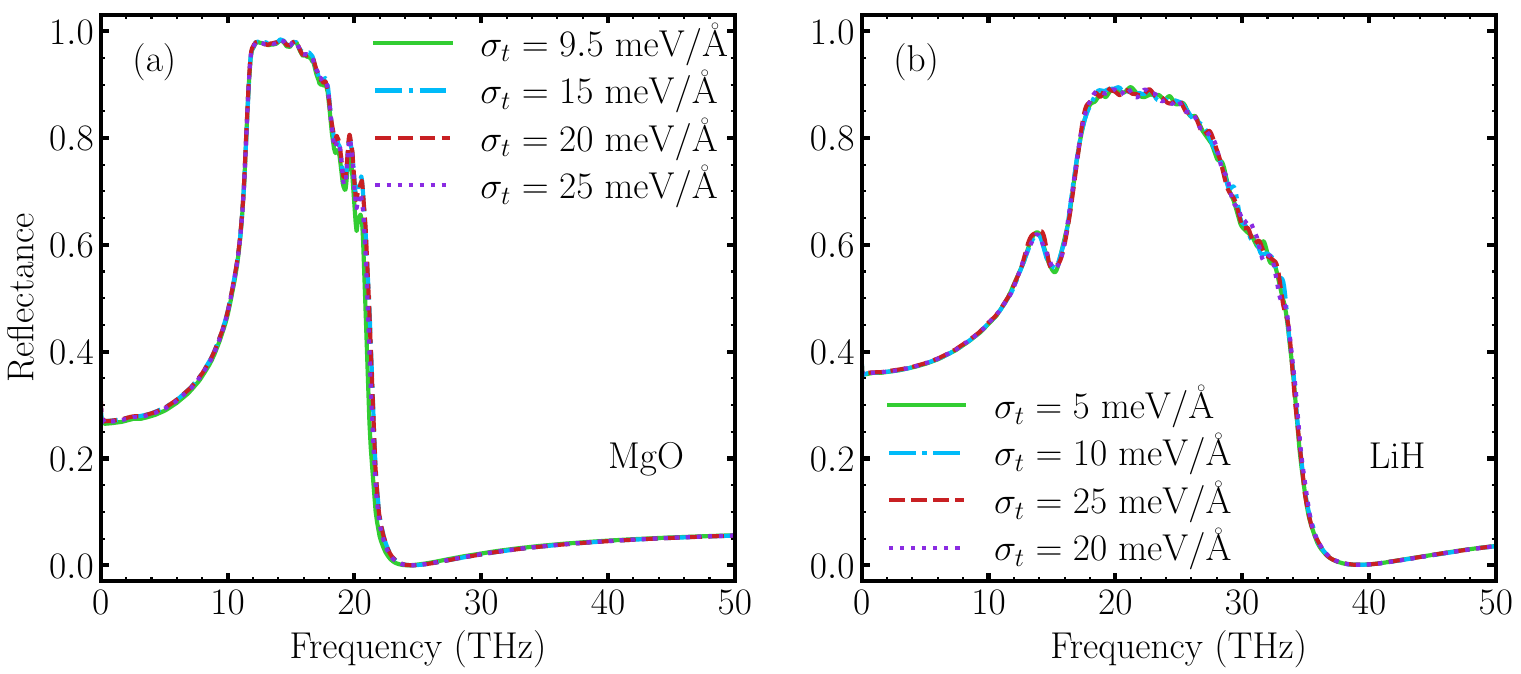}
	\caption{Reflectance of (a)~MgO and (b)~LiH at 300 K calculated from NEP with force errors $\sigma_t$ of different magnitudes.}
	\label{fig:error}
\end{figure}

In order to study the effect of force error in training NEP on the prediction of infrared optical properties, we add additional random forces of different magnitudes in MD at 300 K. In fact, for these materials whose anharmonicity are not very weak, the scattering caused by random forces is much weaker than the intrinsic phonon-phonon scattering. Thus, the effect of force error on the prediction of reflectance is negligible as shown in the Fig.~\ref{fig:error}.
\section{The influence of $\varepsilon_\infty$ and effective charge on reflectance}\label{sec:a3}
The $\varepsilon_\infty$ and Born effective charges calculated by DFPT as well as the experimental $\varepsilon_\infty$ have been described in Sec~\ref{sec2D}. Besides, the experimental effective charge of MgO is 1.96~\cite{PhysRevB.55.R15983}. Experimental effective charge of LiH is obtained from the Szigeti charge $q_s$ and the high-frequency dielectric constant $\varepsilon_\infty$~\cite{LiH1967}, using the relation $q=q_s\left( \varepsilon _{\infty}+2 \right) /3=0.991$. 
\begin{figure}[htbp]
	\centering
	\includegraphics[width=1\linewidth]{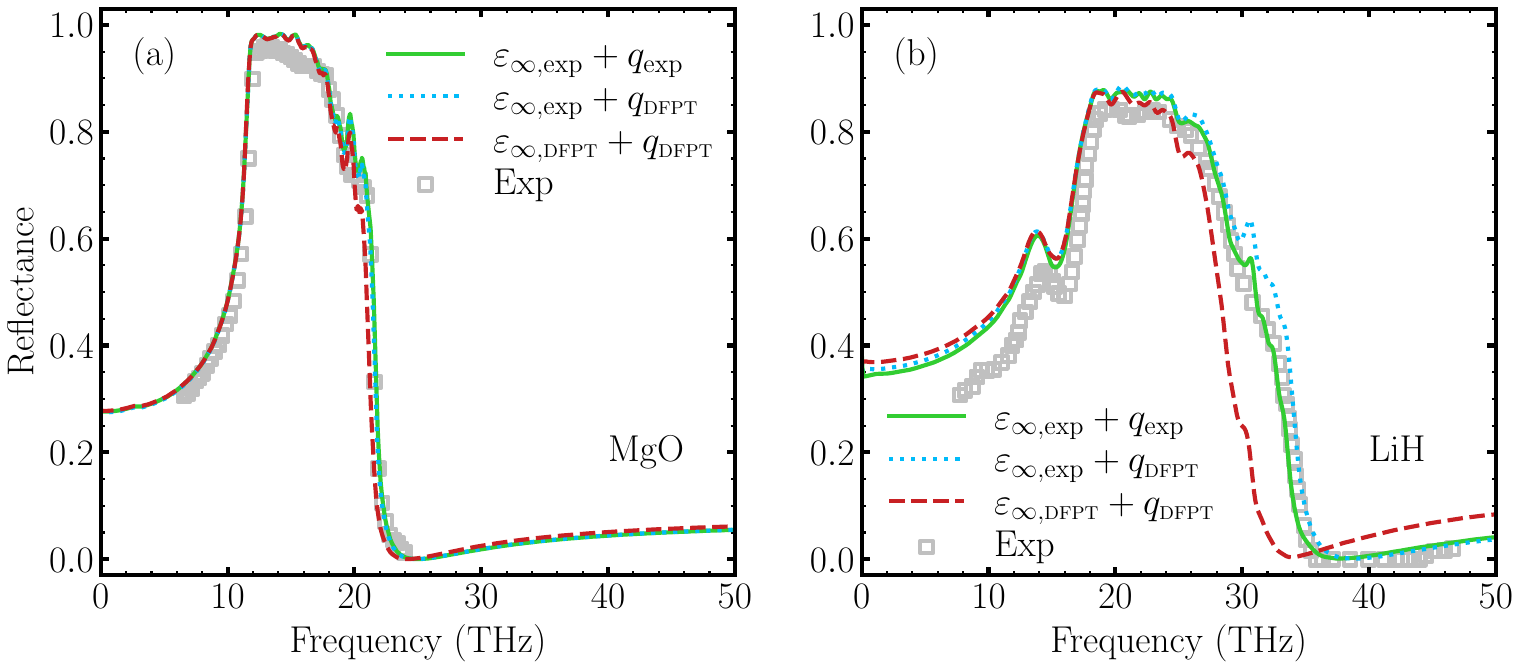}
	\caption{Reflectance of (a)~MgO and (b)~LiH calculated from NEP at 300 K with experimental or DFPT-calculated $\varepsilon_\infty$ and effective charge.}
	\label{fig:nep2}
\end{figure}

As shown in Fig.~\ref{fig:nep2}, the overestimation of $\varepsilon_\infty$ in DFPT leads to an underestimation of $\omega_{\mathrm{LO}}$ and differences between the predicted and experimental reflectance outside the Reststrahlen region. In contrast, the effective charge predicted by DFPT is very close to the experimental value and has negligible effect on the prediction of reflectance.
\section{Phonon self energy of MgO}\label{sec:a4}
\begin{figure}[htbp]
	\centering
	\includegraphics[width=1\linewidth]{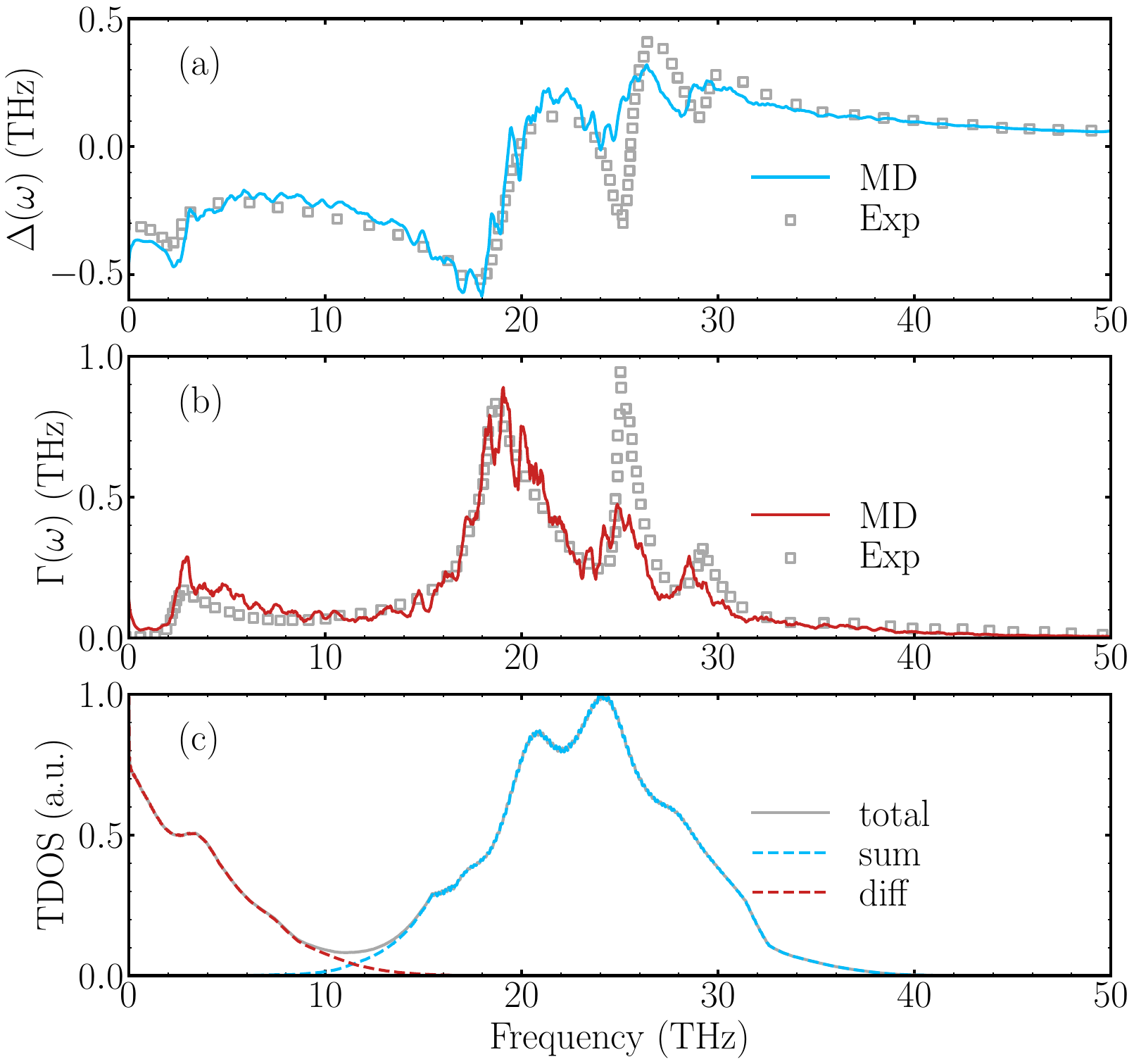}
	\caption{Real~(a) and imaginary part~(b) of phonon self energy of MgO calculated from MD compared with measurement at room temperature~\cite{2004AS}. (c) TDOS of MgO calculated from MD. The blue and red dashed
		lines illustrate the contribution from the phonon summation and difference processes respectively. The full lines
		is the total TDOS.}
	\label{fig:se}
\end{figure}

To assess the accuracy of phonon self-energy derived from MD, we compare our results with experimental fits~\cite{2004AS}, as illustrated in Fig.~\ref{fig:se}. Specifically, given the density of state $\rho_1$ for MgO, we approximate the two phonon density of state~(TDOS) $\rho_2$ following the methodology outlined in~\cite{1975PRB}
\begin{subequations}
\begin{equation}
	\rho _{2,\mathrm{sum}}\left( \omega '' \right) =\int_0^{\infty}{d\omega}\int_0^{\infty}{d\omega '\rho _1\left( \omega ' \right) \rho _1\left( \omega \right) \delta \left( \omega ''-\omega '-\omega \right)},
\end{equation}
\begin{equation}
\rho _{2,\mathrm{diff}}\left( \omega '' \right) =\int_0^{\infty}{d\omega}\int_0^{\infty}{d\omega '\rho _1\left( \omega ' \right) \rho _1\left( \omega \right) \delta \left( \omega ''-\omega '+\omega \right)}.
\end{equation}
\end{subequations}

The phonon self-energy spectrum exhibits three peaks arising from the phonon summation process and a single peak resulting from the phonon difference process. These features correspond to TDOS. In general, the results obtained from MD are in good agreement with experiments. The minor discrepancies observed may be attributed to inaccuracies in the LO phonon frequencies, which stem from the absence of explicit long-range interactions in the model.
\section{Nuclear quantum effect on infrared dielectric function of MgO at 300 K}\label{sec:a5}
\begin{figure}[htbp]
	\centering
	\includegraphics[width=1\linewidth]{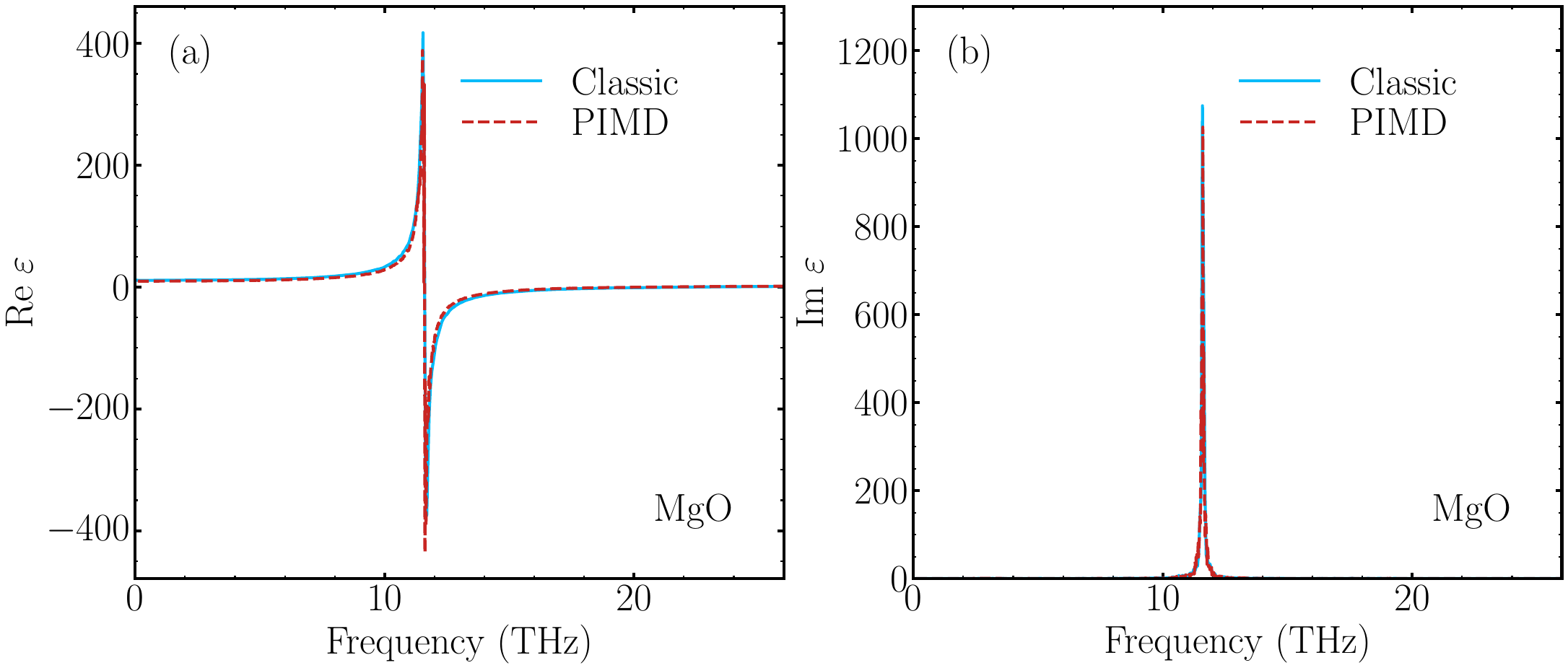}
	\caption{The infrared dielectric function of MgO at 300~K calculated from classical MD~(blue solid lines) and PIMD~(red dashed lines) with NEP: (a) real part, (b) imaginary part.}
	\label{fig:nep3}
\end{figure}

The infrared dielectric function of MgO at 300~K, computed using classical MD and PIMD with NEP potential, is presented in Fig.~\ref{fig:nep3}. At room temperature, the nuclear quantum effects are negligible. However, at cryogenic temperatures, the computational cost of PIMD becomes prohibitively high due to the extremely long phonon lifetimes. This challenging issue will be explored in detail in future work.

\newpage

%
\end{document}